\documentclass[aps,prl,reprint,superscriptaddress]{revtex4-2}

\usepackage{graphicx}
\usepackage[separate-uncertainty=true]{siunitx}
\usepackage{hyperref}
\hypersetup{
	colorlinks=true,
	linkcolor=blue,
	filecolor=blue,      
	urlcolor=blue,
	citecolor=blue,
}

\begin{document}


\title{Energy Compression and Stabilization of Laser-Plasma Accelerators}

\author{A. Ferran Pousa}
\email{angel.ferran.pousa@desy.de}
\affiliation{Deutsches Elektronen-Synchrotron DESY, Notkestr. 85, 22607 
     Hamburg, Germany }

\author{I. Agapov}
\affiliation{Deutsches Elektronen-Synchrotron DESY, Notkestr. 85, 22607 
    Hamburg, Germany }

\author{S. A. Antipov}
\affiliation{Deutsches Elektronen-Synchrotron DESY, Notkestr. 85, 22607 
    Hamburg, Germany }

\author{R. W. Assmann}
\affiliation{Deutsches Elektronen-Synchrotron DESY, Notkestr. 85, 22607 
    Hamburg, Germany }
\affiliation{Laboratori Nazionali di Frascati, Via Enrico Fermi 40, 00044 Frascati, Italy}

\author{R. Brinkmann}
\affiliation{Deutsches Elektronen-Synchrotron DESY, Notkestr. 85, 22607 
    Hamburg, Germany }

\author{S. Jalas}
\affiliation{Department of Physics Universit\"{a}t Hamburg, Luruper Chaussee 149, 22761 Hamburg, Germany}

\author{\\M. Kirchen}
\affiliation{Deutsches Elektronen-Synchrotron DESY, Notkestr. 85, 22607 
    Hamburg, Germany }

\author{W. P. Leemans}
\affiliation{Deutsches Elektronen-Synchrotron DESY, Notkestr. 85, 22607 
    Hamburg, Germany }
\affiliation{Department of Physics Universit\"{a}t Hamburg, Luruper Chaussee 149, 22761 Hamburg, Germany}

\author{A. R. Maier}
\affiliation{Deutsches Elektronen-Synchrotron DESY, Notkestr. 85, 22607 
    Hamburg, Germany }

\author{A. Martinez de la Ossa}
\affiliation{Deutsches Elektronen-Synchrotron DESY, Notkestr. 85, 22607 
    Hamburg, Germany }

\author{J. Osterhoff}
\affiliation{Deutsches Elektronen-Synchrotron DESY, Notkestr. 85, 22607 
    Hamburg, Germany }

\author{M. Th\'{e}venet}
\affiliation{Deutsches Elektronen-Synchrotron DESY, Notkestr. 85, 22607 
    Hamburg, Germany }

\date{\today}

\begin{abstract}
    Laser-plasma accelerators outperform current radiofrequency technology in acceleration strength by orders of magnitude. Yet, enabling them to deliver competitive beam quality for demanding applications, particularly in terms of energy spread and stability, remains a major challenge. In this Letter, we propose to combine bunch decompression and active plasma dechirping for drastically improving the energy profile and stability of beams from laser-plasma accelerators. Realistic start-to-end simulations demonstrate the potential of these post-acceleration phase-space manipulations for simultaneously reducing an initial energy spread and energy jitter of $\SIrange[range-units=single]{\sim1}{2}{\%}$ to ${\lesssim}\SI{0.1}{\%}$, closing the beam-quality gap to conventional acceleration schemes.
\end{abstract}

\maketitle

Laser-plasma accelerators (LPAs)~\cite{PhysRevLett.43.267} can give rise to a new generation of ultra-compact particle accelerators with a wide range of applications. Among others, they could enable cost-effective coherent light sources~\cite{Nakajima2008, Wang2021} or injectors for storage rings~\cite{HILLENBRAND2014153, PhysRevAccelBeams.24.111301}. Improvements in beam quality such as the demonstration of peaked energy spectra~\cite{mangles2004monoenergetic,geddes2004high, faure2004laser}, GeV energy~\cite{Leemans2006, PhysRevLett.113.245002, PhysRevLett.122.084801}, high current~\cite{lundh2011few} and low emittance~\cite{fritzler2004emittance,brunetti2010low, PhysRevSTAB.15.111302}, bring the performance of these devices closer to that of radiofrequency (RF) accelerators. Still, challenges limiting their applicability remain, particularly regarding the beam energy spread and stability.


Applications such as free-electron lasers (FELs) require an energy spread ${\lesssim} \SI{0.1}{\%}$~\cite{RevModPhys.85.1}, yet current LPAs typically operate in the ${\sim} \SI{1}{\%}$ range~\cite{PhysRevX.10.031039}. The main source behind this is typically the steep slope of the accelerating fields, which leads to beams with a strong longitudinal energy correlation (chirp), together with various contributions to the slice energy spread~\cite{PhysRevLett.108.094801, FerranPousa2019,PhysRevAccelBeams.21.111301}. Many techniques have been proposed for reducing the energy chirp of plasma beams, either within the acceleration stage~\cite{PhysRevLett.118.214801, Manahan2017,PhysRevLett.123.054801} or in a dedicated external device~\cite{PhysRevLett.122.034801, PhysRevLett.122.114801, PhysRevLett.122.204804, PhysRevApplied.12.064011, PhysRevLett.112.114801,PhysRevAccelBeams.23.121302,BANE2012106, PhysRevLett.114.114801}. A promising approach is the use of beam loading~\cite{vanderMeer:163918, PhysRevLett.101.145002, doi:10.1063/1.1889444} for flattening the average accelerating gradient along the LPA~\cite{Couperus2017, Kirchen:456053,PhysRevLett.126.104801, PhysRevLett.126.214801}. This has enabled the demonstration of the first sub-percent energy spread beams capable of generating FEL radiation~\cite{Wang2021}.
Nonetheless, reaching the performance of conventional machines demands further improvements to the energy spread as well as to the shot-to-shot energy jitter, which currently ranges in the few percent~\cite{PhysRevX.10.031039,Wang2021}.

The energy stability is critical for the beam transport downstream of the LPA, and thus for virtually any application.
Especially demanding is the injection into diffraction-limited storage rings, where particle energy deviations up to ${\sim} \SI{1}{\%}$~\cite{Schroer:426140} are tolerated.
This requires an energy jitter and energy spread ${\lesssim} \SI{0.1}{\%}$ rms. Recent developments in machine learning and active feedback loops \cite{PhysRevAccelBeams.22.041303,PhysRevLett.126.104801, Shalloo2020} offer a path towards LPAs of improved stability, particularly with the onset of kilohertz lasers~\cite{He2015, doi:10.1063/1.4921159,PhysRevAccelBeams.21.013401}, but a permille energy jitter is yet to be demonstrated.


In this Letter, we propose a technique for drastically -- and simultaneously -- reducing the energy spread and energy jitter of LPA beams in a two-step process.
First, longitudinal decompression in a magnetic chicane is used to imprint a linear correlation between the particles' arrival time and energy~\cite{PhysRevSTAB.11.101301, PhysRevX.2.031019, Mayet:IPAC2017-WEPVA006, Ferran_Pousa_2017,PhysRevLett.123.054801}. Second, a linear longitudinal electric field is applied to remove the imprinted correlation and correct deviations with respect to the target energy. This is carried out by an \textit{active} plasma dechirper (APD), a dedicated plasma stage where the wakefields are generated by a fraction of the LPA driver. In contrast to \textit{passive} plasma dechirpers~\cite{PhysRevLett.122.034801, PhysRevLett.122.114801, PhysRevLett.122.204804, PhysRevApplied.12.064011}, where the wakefields are generated by the electron beam itself, an APD takes advantage of the intrinsic synchronization between the LPA and APD drivers for correcting the central energy jitter, and not only the energy spread. This combination of chicane and APD is the first demonstration of a plasma-based energy compression system~\cite{GILLESPIE1981285, PhysRevSTAB.6.030702}. Its working principle resembles that of chirped-pulse amplification in lasers~\cite{STRICKLAND1985447}. Realistic start-to-end simulations show that this method can be incorporated into state-of-the-art LPAs~\cite{PhysRevLett.126.104801,Kirchen:456053} for improving the energy spread and stability by an order of magnitude, closing the performance gap to RF accelerators.

\begin{figure*}
    \includegraphics[width=\textwidth]{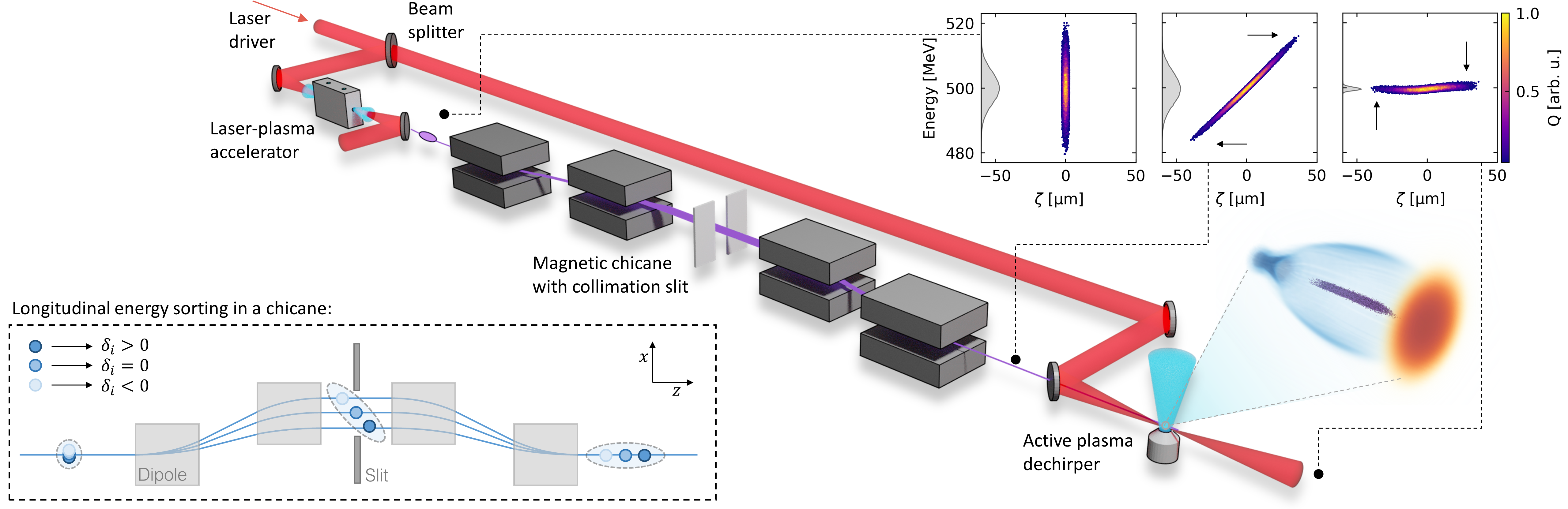}%
    \caption{\label{fig:concept_overview} Basic layout and working principle of an LPA energy compressor and stabilizer. Only the LPA source and the relevant beam line components (laser, chicane and APD) are shown. The longitudinal phase space of the beam at different locations is also displayed, as well as the 3D~\cite{FerranPousa:IPAC2017-TUPIK007} wakefield structure in the APD.}
\end{figure*}

The combined effect of decompression and dechirping can be studied by investigating the single-particle dynamics.
By establishing a reference energy $\gamma_\mathrm{ref}$ as the desired beam energy of the accelerator, a relative energy deviation $\delta(t)=(\gamma(t)-\gamma_\mathrm{ref})/\gamma_\mathrm{ref}$ and longitudinal coordinate $\zeta(t) = z(t) - z_\mathrm{ref}(t)$ can be defined for each particle. Here, $\gamma = \sqrt{1 + (\mathbf{p}/m_ec)^2}$ is the relativistic Lorentz factor, with $\mathbf{p}$ and $m_e$ being, respectively, the momentum and rest mass of an electron; $c$ is the speed of light in vacuum; $t$ is time; $z$ is the longitudinal position; and $z_\mathrm{ref}$ is the position of a reference particle with $\delta=0$ initially located at the beam center.
A dispersive section transforms the phase-space coordinates of a particle initially at ($\zeta_i$, $\delta_i$) to a final position $\zeta_f = \zeta_i + R_{56}\delta_i + \mathcal{O}(\delta_i^2)$~\cite{Chao:2013rba}, where $R_{56}$ is the linear dispersion coefficient, while leaving the energy unchanged.
Thus, to first order in $\delta_i$, a beam with no initial correlation between $\zeta_i$ and $\delta_i$ is longitudinally stretched by a factor
\begin{equation}\label{eq:decompression}
    S \equiv \frac{\sigma_{\zeta_f}}{\sigma_{\zeta_i}} = \sqrt{\left( \frac{R_{56} \sigma_{\delta_i}}{\sigma_{\zeta_i}} \right)^2 + 1} \ ,
\end{equation}
while developing a linear chirp $\chi \equiv \sigma_{\zeta\delta} / \sigma_{\zeta}^{2} = R_{56}^{-1} (1-S^{-2})$, which is $\chi \simeq R_{56}^{-1} $ for $S^2\gg1$. Here, $\sigma_\zeta$ and $\sigma_\delta$ are the standard deviations of $\zeta$ and $\delta$, and $\sigma_{\zeta\delta}$ is their covariance.
After decompression, the beam enters a dechirper of length $L$ that applies a linear longitudinal electric field $E_z(\zeta)=-(m_ec^2/e) \, \mathcal{E}' \, (\zeta - \zeta_0)$ with normalized slope $\mathcal{E}'$ centered at $\zeta_0$, where $e$ is the elementary charge.
Assuming a highly relativistic beam ($\gamma \gg 1$), $\zeta_f$ stays constant throughout the dechirper and the particle energy is transformed into a final value
\begin{equation}\label{eq:delta}
    \delta_f = \frac{1}{R_{56}}(\zeta_f-\zeta_i) + \frac{\mathcal{E}'L}{\gamma_\mathrm{ref}}(\zeta_f - \zeta_0)\ .
\end{equation}

Therefore, the energy correlation imprinted by the linear dispersion can be removed by the dechirper if
\begin{equation}\label{eq:dechirp_cond}
    \mathcal{E}'L = -\frac{ \gamma_{\rm ref}}{R_{56}} \ .
\end{equation}
This results in a net reduction of the beam energy spread, whose final value is fully determined by $R_{56}$ as
\begin{equation} \label{eq:final_spread}
    \sigma_{\delta_f} = \frac{\sigma_{{\zeta}_i}}{R_{56}} \simeq \frac{\sigma_{\delta_i}}{S}\ ,
\end{equation}
where the last equality holds if $S^2 \gg 1$.

This technique is ideally suited for LPA beams.
As Eq. (\ref{eq:final_spread}) shows, the typically ultra-short (\SI{\sim 1}{\micro\metre}) length and large (\SI{\sim 1}{\%}) energy spread allow for a factor 10 decompression and energy spread reduction with minimal dispersion ($R_{56}\sim \SI{1}{\milli\metre}$).
In addition, the high initial peak current (up to \SI{{\sim}10}{\kilo\ampere} \cite{lundh2011few}) means that a final current in the $\SI{\sim 1}{\kilo\ampere}$ range can still be achieved after decompression, allowing for FEL applications.
When high current is not required, such as in storage ring injectors, an even more drastic energy spread reduction could be realized.

As illustrated in Fig. \ref{fig:concept_overview}, the bunch decompression can be performed by a magnetic chicane, where path length differences arise due to an energy-dependent transverse deflection. This results in $R_{56}=2 \theta^2(L_d+2L_m/3)$~\cite{Chao:2013rba}, where $L_m$ and $\theta$ are, respectively, the magnet length and bending angle (for $\delta=0$); and $L_d$ is the distance between the central and outer dipoles.

When Eq. (\ref{eq:dechirp_cond}) is satisfied, Eq. (\ref{eq:delta}) also yields that the final deviation of the average beam energy is given by 
\begin{equation}\label{eq:final_gamma}
    \langle \delta_f \rangle = \frac{\zeta_0}{R_{56}} \ .
\end{equation}
Therefore, if $\zeta_0=0$, the final beam energy is stabilized to $\gamma_\mathrm{ref}$ regardless of its initial value. This requires the ability to control $\zeta_0$ independently of the beam position, i.e., with an \textit{active} dechirping medium where the fields are not generated by the beam itself. An APD accomplishes this in a compact, plasma-based setup.
It is conceptually similar to a laser-plasma lens~\cite{PhysRevSTAB.17.121301,Thaury2015}, but aimed at the generation of longitudinal, instead of transverse, fields with a fraction of the LPA driver.
The intrinsic synchronization between the LPA and APD drivers allows for a precise control of $\zeta_0$, independently  of the  electron beam arrival time. This setup is also robust against realistic timing jitters between both drivers.
As obtained from Eq. (\ref{eq:final_gamma}), if ${R_{56}\sim\SI{1}{\milli\metre}}$, a state-of-the-art timing jitter of ${\lesssim} \SI{10}{\femto\second}$~\cite{Schulz2015, shalloo2015measurement} is sufficient for achieving a per-mille energy jitter.

\begin{figure}
    \includegraphics[width=\columnwidth]{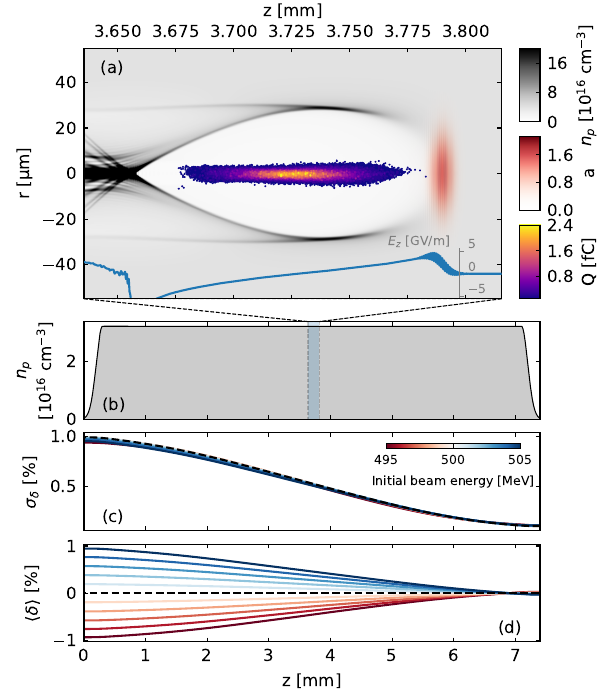}%
    \caption{\label{fig:apd_evolution} (a) Plasma wakefields and electron beam at the center of the APD. (b) APD density profile. (c) Energy spread  and (d) average energy deviation along the APD of beams with initial energy deviations between \SI{\pm1}{\%}. The black dashed line corresponds to a reference initial energy of \SI{500}{\mega\electronvolt}.}
\end{figure}

When the peak normalized vector potential of the APD driver is sufficiently high (i.e, $a_0 \gtrsim 2$), large plasma electron cavitation occurs and a trailing wakefield with uniform $\mathcal{E}'$ is generated (cf. Fig \ref{fig:apd_evolution}(a)). The length of the cavity is approximately given by the plasma wavelength $\lambda_p=2\pi/k_p$, where $k_p = (n_pe^2/m_e\epsilon_0c^2)^{1/2}$ and $n_p$ are the plasma electron wavenumber and density and $\epsilon_0$ is the vacuum permittivity.
As depicted in Fig. \ref{fig:concept_overview}, a slit in the center of the chicane removes particles beyond a maximum, $\delta_\mathrm{max}$, and minimum, $\delta_\mathrm{min}$, energy deviation for ensuring that the stretched beam fits within the cavity.
Imposing a total beam extension ${\lesssim}\lambda_p/2$ yields the condition $\lambda_p \gtrsim 2\,(\delta_\mathrm{max}-\delta_\mathrm{min})R_{56}=2\,(\delta_\mathrm{max}-\delta_\mathrm{min})\sigma_{\zeta_i}/\sigma_{\delta_f}$, which determines the maximum plasma density for achieving a certain final energy spread.
For $\sigma_{\delta_f}=\num{e-3}$, $\sigma_{\zeta_i}=\SI{1}{\micro\metre}$ and $\delta_\mathrm{max}-\delta_\mathrm{min}=0.06$, $n_p\lesssim\SI{8e16}{\per\cubic\centi\metre}$ is obtained.
The field slope $\mathcal{E}'$ can be estimated from the non-linear cold fluid equation \cite{RevModPhys.81.1229} for the wakefield potential, $\psi$, behind the driver, i.e., $\mathcal{E}'(\zeta) = \partial_\zeta^2 \psi(\zeta) = - k_p^2\,(1 - 1/(1+\psi(\zeta))^2) / 2$.
At $\zeta_0$, which occurs around the center of the cavity, $\psi$ is maximum and given approximately by $\psi_0\equiv\psi(\zeta_0) \sim \hat{w}_0^2/4$~\cite{doi:10.1063/1.2203364}, where $\hat{w}_0=k_pw_0$ and $w_0$ is the spot size of the laser at focus. This implies that $\mathcal{E}'(\zeta_0) \sim - k_p^2\,(1 - 1/(1+\hat{w}_0^2/4)^2) / 2$. Coupled with Eq. (\ref{eq:dechirp_cond}), this expression allows for an estimate of the required APD length, under the assumption that $w\sim w_0$ throughout the dechirper.
Relative to the laser Rayleigh length, $Z_R=\pi w_0^2/\lambda_0$ \cite{RevModPhys.81.1229}, the APD length is found to be $L/Z_R = 2\gamma_\mathrm{ref}\lambda_0\,(4 + \hat{w}_0^2)^2/(\pi R_{56}\hat{w}_0^4 (8 + \hat{w}_0^2))$, where $\lambda_0$ is the laser wavelength.
For $R_{56}=\SI{1}{\milli\metre}$, $\gamma_\mathrm{ref}=\num{e3}$ and $\lambda_0=\SI{800}{\nano\metre}$, this expression yields $\hat{w}_0 \gtrsim 1$ for ensuring $L\lesssim Z_R$ (i.e. $w\sim w_0$).
Under this condition, a compact, \si{\milli\metre}-long APD can be realized without external laser guiding.
Given the typically low density and narrow driver, no self-injection and, thus, no dark current is expected from the APD~\cite{doi:10.1063/1.2173960,doi:10.1063/1.1799371}.

The performance of the technique is demonstrated by means of two comprehensive simulation studies of an energy compression system.
First, the setup is probed with an ideal Gaussian electron bunch to generally assess the energy spread and jitter correction.
Second, a full start-to-end study including a realistic LPA and relevant experimental jitters validates its efficacy under real-world conditions.
The initial beam capture and final focus into the APD are carried out by active plasma lenses~\cite{PhysRevLett.115.184802}.
This enables a compact setup with minimal chromatic emittance growth~\cite{PhysRevSTAB.16.011302}, but other options are also possible~\cite{PhysRevAccelBeams.24.014801,PhysRevAccelBeams.24.111301}.
After initial prototyping with \texttt{Wake-T}~\cite{Ferran_Pousa_2019}, the plasma elements are simulated with the quasi-3D particle-in-cell code \texttt{FBPIC}~\cite{LEHE201666} and the conventional elements with \texttt{Ocelot}~\cite{AGAPOV2014151}, including the effects of 3D space charge and 1D coherent synchrotron radiation (CSR). 
Using \texttt{libEnsemble}~\cite{libEnsemble2022}, the jitter of the setup is comprehensively evaluated with hundreds of simulations.
See \cite{SupplementalMaterial} for additional simulation details.

The parameters of the probe Gaussian electron beam are representative of current state-of-the-art LPAs~\cite{PhysRevLett.126.104801, Kirchen:456053,PhysRevLett.126.214801,Wang2021}, having a \SI{500}{\mega\electronvolt} energy with \SI{1}{\%} rms shot-to-shot jitter, \SI{1}{\%} rms energy spread, \SI{1}{\micro\metre} normalized emittance, \SI{2}{\micro\metre} transverse size, \SI{0.5}{\milli\radian} rms divergence, \SI{10}{\femto\second} FWHM duration, and \SI{10}{\pico\coulomb} charge.
The chicane has a total length of \SI{2}{\metre}, with $L_d=\SI{50}{\centi\metre}$,  $L_m=\SI{20}{\centi\metre}$ and $\theta=\SI{34.4}{\milli\radian}$, resulting in $R_{56} = \SI{1.5}{\milli\metre}$ and $S=11.8$. It includes a collimating slit with a \SI{1.4}{\milli\metre} horizontal aperture for filtering particles with $|\delta|> \SI{3}{\%}$.
The APD has a \SI{6.8}{\milli\metre} plateau with a \SI{3.2e16}{\per\cubic\centi\metre}  density and two \SI{0.3}{\milli\metre} Gaussian ramps at the entrance and exit.
The APD driver is a \SI{2}{\joule} Gaussian laser pulse focused at the center of the plateau with $a_0=2.15$, $\lambda_0=\SI{0.8}{\micro\metre}$, $w_0=\SI{22}{\micro\metre}$, and a \SI{25}{\femto\second} FWHM duration.
The plasma lenses have a \SI{1}{\centi\metre} length, \SI{1.62}{\kilo\tesla\per\metre} focusing gradient, and \SI{e15}{\per\cubic\centi\metre} density.
They are placed \SI{10}{\centi\metre} down and upstream of the initial beam and the APD, respectively.

\begin{figure}
    \includegraphics[width=\columnwidth]{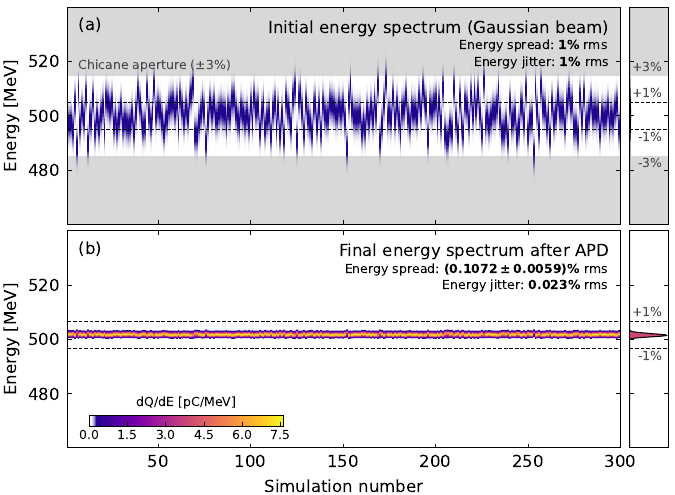}%
    \caption{\label{fig:waterfall_plot} 
    (a) Initial energy spectrum of 300 Gaussian beams with a \SI{1}{\%} rms energy jitter and energy spread. The gray area represents the energies filtered out by the slit in the chicane. (b) Final energy spectrum after the APD.
    }
\end{figure}

Fig. \ref{fig:apd_evolution} shows the evolution of the decompressed Gaussian beam within the APD for initial energy deviations between $\pm\SI{1}{\%}$. The wakefields generated by the driver effectively reduce the energy spread while simultaneously correcting the initial energy deviations. The reference beam has a final energy spread of ${\sim}\SI{0.10}{\%}$ (total) and ${\sim}\SI{0.084}{\%}$ (slice average).
This agrees with Eq. (\ref{eq:final_spread}), which predicts a value of ${\sim}\SI{0.085}{\%}$.
The total energy spread is larger due to non-linearities in $E_z$ arising mostly from beam loading.
The efficacy of the energy compression can be clearly seen in Fig. \ref{fig:waterfall_plot}. The results from 300 simulations show that the initial energy jitter of \SI{1}{\%} is reduced to \SI{0.023}{\%}. Similarly, the initial energy spread of \SI{1}{\%} is reduced by a factor ${\sim}10$ to $\SI{0.1072\pm0.0059}{\%}$.
The final normalized emittances of $\SI{1.22\pm0.13}{\micro\metre}$ (horizontal) and $\SI{1.179\pm0.086}{\micro\metre}$ (vertical) show only a slight increase dominated by chromatic effects during capture and focus. These values correspond to the average and rms deviations of all simulated shots.
Ultimately, this study demonstrates that the energy compressor behaves as expected from theory, improving the energy spread and stability by at least an order of magnitude.

\begin{figure}
    \includegraphics[width=\columnwidth]{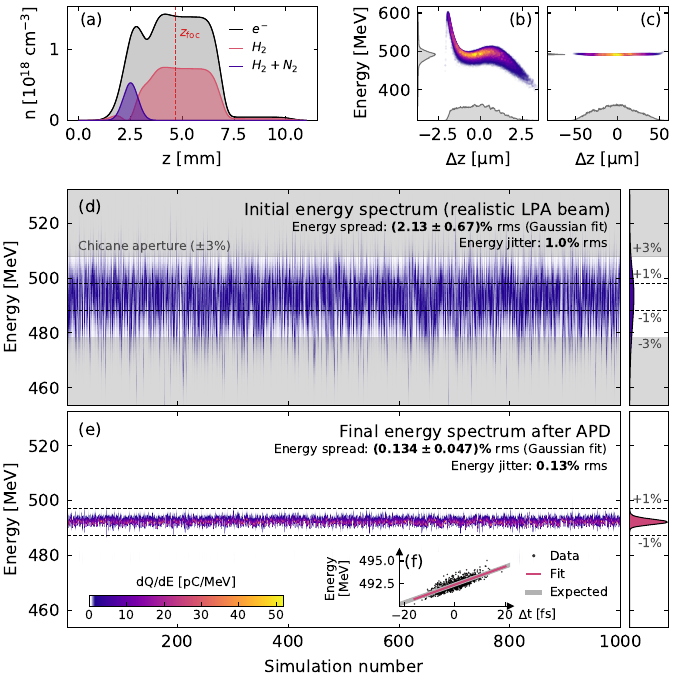}%
    \caption{\label{fig:waterfall_plot_s2e} Results from the start-to-end jitter study: (a) LPA density profile; longitudinal phase space of the reference beam (i.e., no jitters) at the exit of the (b) LPA and (c) APD; beam energy spectra at the (d) LPA and (e) APD exits; (f) average beam energy versus laser arrival time jitter ($\Delta t$), including the expected correlation and a linear fit to the data.}
\end{figure}

The real-world applicability of the energy compressor is validated through a full start-to-end study including a realistic LPA as well as relevant experimental jitters.
The LPA used for this study is based on downramp-assisted ionization injection, a well-proven technique which can be accurately simulated~\cite{Kirchen:456053, PhysRevLett.126.104801} and where the dominant sources of jitter are well known: laser focal position, energy, and pulse duration~\cite{PhysRevX.10.031039,Kirchen:456053}.
It is designed as an evolution of the LUX target \cite{PhysRevX.10.031039,Kirchen:456053, PhysRevLett.126.104801} aimed at generating \SI{500}{\MeV} beams with maximum stability to laser jitters.
The density profile, displayed in Fig. \ref{fig:waterfall_plot_s2e}(a), contains a mixture of $\mathrm{H_2}$ and $\mathrm{N_2}$ (\SI{1}{\%}) for electron injection, a \SI{1.46e18}{\per\cubic\centi\metre} plateau for acceleration, and a low-density tail (\SI{4e16}{\per\cubic\centi\metre}) for divergence minimization~\cite{SupplementalMaterial}.
The laser driver is a \SI{130}{\tera\watt} Ti:Sa system with a total energy of \SI{4.68}{\joule}, split between the LPA (\SI{2.68}{\joule}) and the APD (\SI{2}{\joule}).
Its longitudinal profile is Gaussian with a FWHM duration of \SI{34}{\femto\second}, while its transverse profile is modeled as a so-called flattened Gaussian~\cite{doi:10.1080/09500349708232927}.
This consists of a sum of Laguerre-Gauss modes that accurately describes flat-top high-power lasers in experiments~\cite{kirchen2021novel}.
The laser is subject to realistic jitters in the focal plane position (\SI{100}{\micro\metre} rms), energy (\SI{0.5}{\%} rms) and pulse duration (\SI{1}{\%} rms)~\cite{Kirchen:456053,WU2020106453}.
Transverse and longitudinal (i.e., timing) jitters between the LPA and APD pulses of \SI{5}{\micro\metre}~\cite{kirchen2021novel} and \SI{5}{\femto\second} rms, respectively, are also included at the APD entrance.
The LPA driver is focused to $w_0=\SI{21}{\micro\metre}$, with $a_0=2.21$, at $z_\mathrm{foc}=\SI{4.68}{\milli\metre}$ into the target.
The data from 1000 simulations shows that the resulting LPA beams have an energy of \SI{494.3\pm4.9}{\mega\electronvolt} (i.e., \SI{1}{\%} jitter), an rms (Gaussian fit) energy spread  of $\SI{2.13\pm0.67}{\%}$, a normalized emittance of $\SI{2.48\pm0.12}{\micro\metre}$ (horizontal) and $\SI{0.749\pm0.062}{\micro\metre}$ (vertical), a divergence of \SI{0.762\pm0.026}{\milli\radian} (horizontal) and \SI{0.350\pm0.029}{\milli\radian} (vertical), a charge of $\SI{49.8\pm5.6}{\pico\coulomb}$, a FWHM duration of $\SI{8.96\pm0.57}{\femto\second}$ and a peak current of $\SI{5.63\pm0.87}{\kilo\ampere}$.
A realistic pointing jitter of $\SI{0.5}{\milli\radian}$, consistent with experiments~\cite{Kirchen:456053}, is externally added.
To transport this beam, the focusing gradient in the first and second plasma lenses is tuned to \SI{1.60}{\kilo\tesla\per\metre} and \SI{2.37}{\kilo\tesla\per\metre}, respectively.
The APD is placed \SI{6.6}{\centi\metre} downstream of the second lens and has a \SI{5.4}{\milli\metre} plateau with a \SI{4.1e16}{\per\cubic\centi\metre} density.
The laser driver is focused at the center of the APD with $w_0=\SI{27.5}{\micro\metre}$ and $a_0=1.48$.

The results of this jitter scan can be seen in Fig. \ref{fig:waterfall_plot_s2e}. After the APD, the beams have an average energy of \SI{492.41\pm0.63}{\mega\electronvolt} (i.e., \SI{0.13}{\%} jitter) and an rms (Gaussian fit) energy spread of $\SI{0.134\pm0.047}{\%}$ (total) and $\SI{0.071\pm0.012}{\%}$ (slice).
This is an order of magnitude improvement over the initial values, and demonstrates a dechirping strength of \SI{{\sim} 62}{\giga\electronvolt\per\milli\metre\per\metre}, a factor ${>} \num{e3}$ higher than with RF technology~\cite{PhysRevAccelBeams.24.111301}.
The final energy variability is dominated by the timing jitter between the two laser pulses. This can be seen in Fig.~\ref{fig:waterfall_plot_s2e}(f), where the observed time-energy correlation is in full agreement with Eq.~(\ref{eq:final_gamma}).
The final beam emittances of $\SI{5.4\pm1.2}{\micro\metre}$ (horizontal) and $\SI{1.78\pm0.85}{\micro\metre}$ (vertical) experience an increase mostly due to chromatic effects in the transport line and transverse offsets of the laser at the APD entrance, which also lead to an increased pointing jitter of $\SI{1.91}{\milli\radian}$ (horizontal) and $\SI{1.84}{\milli\radian}$ (vertical).
The final beam charge of $\SI{34.8\pm5.6}{\pico\coulomb}$ is reduced due to the collimating slit in the chicane, resulting in a peak current of $\SI{0.180\pm0.038}{\kilo\ampere}$ for a bunch duration of $\SI{193\pm46}{\femto\second}$.
This study demonstrates the feasibility and robustness of the proposed concept under real-world conditions, paving the way towards the experimental demonstration of reliable and high-quality plasma accelerators.
Side effects such as emittance growth or charge loss can be greatly minimized by a lower initial energy spread, and the final energy stability can be further improved if the laser timing jitter is reduced.

In conclusion, the presented concept of bunch decompression and active plasma dechirping effectively corrects the energy spread and jitter of LPAs in a compact setup. Large-scale realistic start-to-end simulations demonstrate that the beam energy spread and energy jitter of state-of-the-art LPAs can be reduced by an order of magnitude to the permille and sub-permille range. This would enable LPAs as compact beam sources for future storage rings or free-electron lasers.

\begin{acknowledgments}
    We thank R. Lehe for providing access to the optimization library used to fine-tune the APD parameters in the presented studies.
    This research was supported in part through the Maxwell computational resources operated at Deutsches Elektronen-Synchrotron DESY, Hamburg, Germany. The authors gratefully acknowledge the Gauss Centre for Supercomputing e.V. (www.gauss-centre.eu) for funding this project by providing computing time through the John von Neumann Institute for Computing (NIC) on the GCS Supercomputer JUWELS~\cite{JUWELS} at J\"{u}lich Supercomputing Centre (JSC).
\end{acknowledgments}

\bibliography{references}

\end{document}